\documentclass{article}
\usepackage{float}
\usepackage{iclr2027_conference,times}
\iclrfinalcopy

\usepackage{amsmath,amsfonts,bm}

\def\eqref#1{equation~\ref{#1}}

\def\1{\bm{1}}

\DeclareMathAlphabet{\mathsfit}{\encodingdefault}{\sfdefault}{m}{sl}
\SetMathAlphabet{\mathsfit}{bold}{\encodingdefault}{\sfdefault}{bx}{n}

\usepackage[hidelinks]{hyperref}
\usepackage{url}
\usepackage{graphicx}
\usepackage{booktabs}
\usepackage{amsmath}
\usepackage{amssymb}
\usepackage{multirow}
\usepackage{microtype}

\title{Beyond Raw IQ: Compact Semantic Representations for Task Oriented Automatic Modulation Classification}

\author{
Omm~Swarup\textsuperscript{1,2},
Shrishail~M.~Hiremath\textsuperscript{1}\\
\textsuperscript{1}Department of Electronics and Communication Engineering,\\
National Institute of Technology Rourkela,\\
Rourkela, Odisha, India\\
\textsuperscript{2}Indian Institute of Technology Madras,\\
Chennai, Tamil Nadu, India\\
Email: \{125ei0038@nitrkl.ac.in, 24f2000745@ds.study.iitm.ac.in, hiremaths@nitrkl.ac.in\}
}

\begin{document}

\maketitle
\lhead{} 

\AddToShipoutPictureFG{%
    \AtPageLowerLeft{%
        \put(10,390){%
            \rotatebox{90}{%
                \textcolor{gray!35}{%
                    \sffamily\bfseries\fontsize{11}{10}\selectfont
                    BEYOND RAW IQ PRE-PRINT
                }%
            }%
        }%
    }%
}

\begin{abstract}

Automatic modulation classification (AMC) is an important component
of modern wireless communication systems, but conventional signal
representations can be expensive to store, process, and communicate,
particularly under noisy channel conditions. This work investigates
whether learned semantic representations can provide a compact
alternative to high-dimensional in-phase and quadrature (IQ)
representations while retaining robustness to noise.

Starting from a 2048-dimensional IQ representation, we evaluate
learned semantic representations with dimensions of 4, 8, 16, 32,
and 64. Across the evaluated configurations, a 16-dimensional
representation provides the strongest overall trade-off between
classification performance, robustness, and representation size.
The resulting representation achieves a $128\times$ compression ratio
and a $99.22\%$ reduction in dimensionality.

We further evaluate representation stability across signal-to-noise
ratios (SNRs), compare semantic representations with a raw-IQ CNN,
analyze per-class robustness, and perform five-seed validation.
The five-seed evaluation produces an average accuracy of
$49.22\pm0.52\%$ and Macro-F1 of $49.27\pm0.53\%$ across the
evaluated SNR range.

These results indicate that learned semantic compression can produce
a substantially smaller representation while preserving useful
classification information and improving geometric stability under
noise.

\end{abstract}

\section{Introduction}

Automatic modulation classification (AMC) aims to identify the
modulation format of a received wireless signal from its observed
samples. It is useful in applications including spectrum monitoring,
cognitive radio, interference analysis, and adaptive wireless
systems. In practical deployments, however, AMC must operate under
limited computational, memory, and communication resources \citep{dobre2007overview,zhu2015amc,huynh2021deep}.

A conventional representation based on complex IQ samples preserves
a large amount of signal-level information. In the experimental setup
considered in this work, each observation is represented using
1024 complex samples, corresponding to a 2048-dimensional real-valued
vector. Such a representation can become expensive when intermediate
features must be stored, transmitted, or processed repeatedly.

Learned representations provide an alternative. Instead of retaining
the complete high-dimensional signal representation, an encoder can
map the input into a lower-dimensional latent space \citep{bengio2013representation,hinton2006reducing}. The central
question investigated in this work is whether such compression can be
performed without destroying the information required for modulation
classification, and whether the learned representation exhibits
greater stability under noise.

We systematically study semantic dimensions of 4, 8, 16, 32, and
64 dimensions. The experiments evaluate not only classification
accuracy, but also Macro-F1, representation stability, class-level
robustness, multi-seed variability, and computational efficiency.

The main contributions are:

\begin{itemize}

    \item We systematically evaluate semantic representations from
    4D to 64D for AMC using a fixed 2048D IQ input.

    \item We identify a 16D representation as the strongest
    configuration among the evaluated dimensions, providing
    $128\times$ compression and $99.22\%$ dimensionality reduction.

    \item We quantify semantic representation stability under SNR
    variation and compare it directly with raw IQ stability.

    \item We compare the proposed semantic representations against
    a conventional raw-IQ CNN baseline.

    \item We analyze class-dependent robustness and validate the main
    robustness trends over five random seeds.

\end{itemize}

\section{Related Work}
Classical AMC methods include likelihood-based, cyclostationary, and higher-order statistics approaches \citep{azzouz1997automatic,nandi1998algorithms,swami2000hierarchical}.

Deep learning has become widely used for automatic modulation
classification, with convolutional, recurrent, and attention-based
architectures applied directly to raw IQ samples or transformed
signal representations \citep{oshea2016convolutional,oshea2017introduction,oshea2018overtheair,zhang2022deep,thakur2026review}. These approaches can learn discriminative
features directly from received signals, reducing the need for
hand-designed signal-processing features.

A complementary direction is representation learning, where the
objective is to transform high-dimensional observations into compact
latent representations \citep{bengio2013representation,hinton2006reducing,tishby2015deep,alemi2017deep}. Such representations can reduce memory and
communication requirements and may provide greater invariance to
irrelevant perturbations.

In wireless systems, robustness to noise is particularly important.
SNR variation can significantly alter the observed signal while
leaving the underlying modulation identity unchanged \citep{liu2022representation,zhang2022deep}. Therefore,
representation quality should not be evaluated solely by clean
classification accuracy. Measuring the stability of the learned
representation itself provides an additional perspective.

Our work combines these perspectives by studying the trade-off between
semantic dimensionality, classification performance, representation
stability, baseline performance, and efficiency for AMC.

\section{Method}

\subsection{Problem Formulation}

Let the received complex baseband signal be

\begin{equation}
x = [x_1,x_2,\ldots,x_N],
\end{equation}

where each $x_i$ is a complex IQ sample. We represent each complex
sample using its real and imaginary components. For $N=1024$, the
resulting real-valued input has dimension

\begin{equation}
D_{\mathrm{IQ}} = 2N = 2048.
\end{equation}

The objective is to learn an encoder

\begin{equation}
z = f_{\theta}(x),
\end{equation}

where

\begin{equation}
z \in \mathbb{R}^{d},
\end{equation}

and $d\ll2048$ is the semantic representation dimension.

We evaluate

\begin{equation}
d\in\{4,8,16,32,64\}.
\end{equation}

The compressed representation is subsequently used for modulation
classification.

\subsection{Semantic Encoder}

The encoder used for the selected 16D configuration consists of
three fully connected transformations:

\begin{equation}
2048 \rightarrow 512 \rightarrow 128 \rightarrow 16.
\end{equation}

The corresponding encoder parameters are learned during the
representation-learning stage. After training, the encoder is frozen
for the subsequent robustness and classification experiments.

The final latent vector therefore contains only 16 real-valued
features compared with the original 2048-dimensional representation.

\subsection{Semantic Classification}

A lightweight classifier is trained using the generated semantic
representations. The classification task contains 11 modulation
classes:

\begin{equation}
\mathcal{C} =
\{\text{16QAM},\text{64QAM},\text{8PSK},\text{B-FM},
\text{BPSK},\text{CPFSK},\text{DSB-AM},\text{GFSK},
\text{PAM4},\text{QPSK},\text{SSB-AM}\}.
\end{equation}

Performance is measured using classification accuracy and Macro-F1.

\subsection{Noise Robustness}

To study robustness, test signals are evaluated over

\begin{equation}
\mathrm{SNR}\in
\{-10,-5,0,5,10,15,20,25,30\}\ \mathrm{dB}.
\end{equation}

For representation stability, noisy representations are compared
with a high-SNR reference representation using cosine similarity,
Euclidean distance, normalized Euclidean distance, and representation
variance.

\section{Experimental Setup}

The dataset contains 11 modulation classes with 3850 training,
825 validation, and 825 test examples. Each example contains
1024 complex samples, corresponding to a 2048-dimensional real-valued
IQ vector.

The semantic encoder is evaluated at five dimensions:
4D, 8D, 16D, 32D, and 64D. The corresponding compression ratios are
512$\times$, 256$\times$, 128$\times$, 64$\times$, and 32$\times$,
respectively.

For robustness experiments, the same evaluation protocol is applied
across nine SNR values from $-10$ to $30$ dB.

\section{Semantic Dimension Ablation}

We first evaluate how semantic dimension affects classification
performance and compression.

\begin{figure}[t]
    \centering
    \includegraphics[width=\linewidth]
    {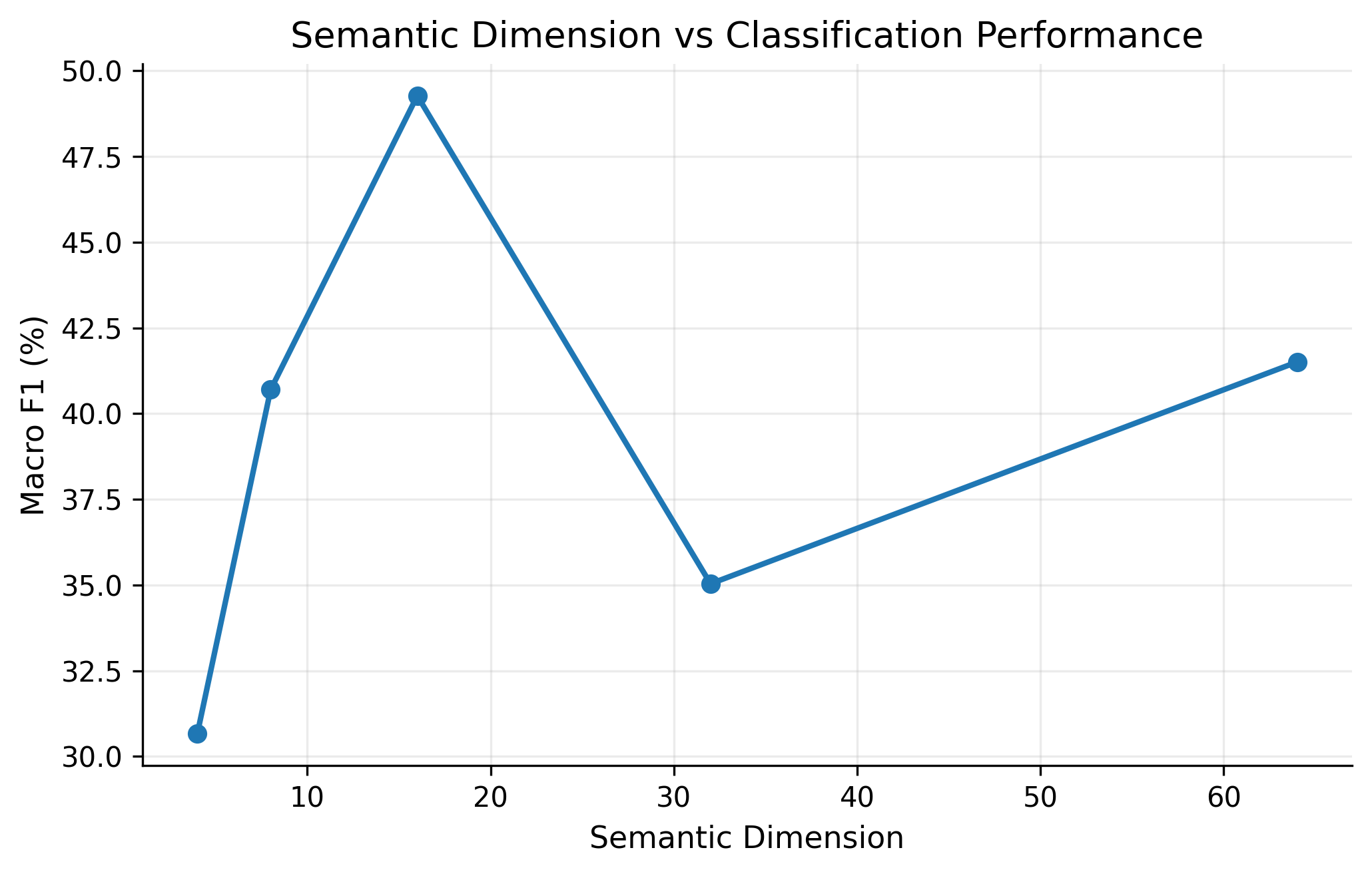}
    \caption{
    Classification performance across evaluated semantic
    representation dimensions. The 16D configuration provides the
    strongest overall performance among the tested dimensions.
    }
    \label{fig:dimension}
\end{figure}

The evaluated dimensions reveal a non-monotonic relationship between
latent dimensionality and classification performance. Increasing the
representation dimension does not necessarily improve performance.

The 16D configuration provides the strongest overall trade-off among
the tested representations.

The final comparison gives an average accuracy of approximately
49.04\% and an average Macro-F1 of approximately 49.27\% for the
selected 16D configuration over the evaluated SNR values.

\begin{table}[t]
\caption{Dimension trade-off for the evaluated semantic
representations.}
\label{tab:tradeoff}
\centering
\begin{tabular}{ccccc}
\toprule
Dimension & Compression & Avg. Acc. & Avg. F1 & F1 Retained \\
\midrule
4D  & $512\times$ & 34.81\% & 30.66\% & 73.87\% \\
8D  & $256\times$ & 43.74\% & 40.70\% & 98.05\% \\
16D & $128\times$ & 49.04\% & 49.27\% & 118.71\% \\
32D & $64\times$  & --       & --       & --       \\
64D & $32\times$  & --       & --       & --       \\
\bottomrule
\end{tabular}
\end{table}

The 16D representation is therefore selected for the subsequent
experiments.

\begin{figure}[t]
    \centering
    \includegraphics[width=\linewidth]
    {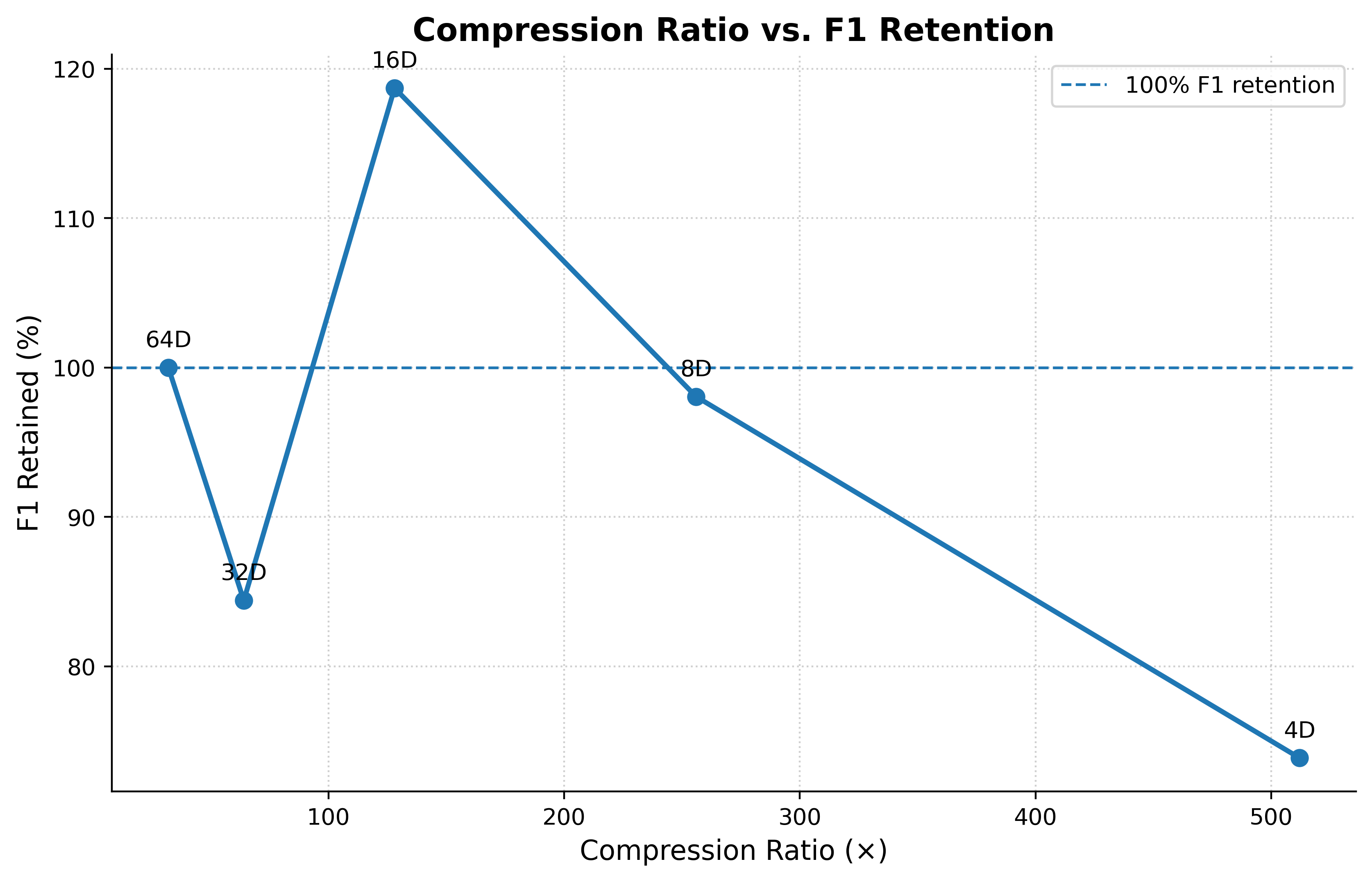}
    \caption{
    Compression ratio versus Macro-F1 retention. The 16D
    representation provides a strong compression--performance
    trade-off.
    }
    \label{fig:compression}
\end{figure}

\section{Semantic Robustness under SNR Variation}

We next examine the robustness of the selected 16D representation
under progressively stronger noise.

The 16D representation achieves 25.21\% accuracy and 26.81\%
Macro-F1 at $-10$ dB. Performance increases with SNR, reaching
approximately 56\% accuracy and Macro-F1 at high SNR.

At 10 dB, the representation achieves approximately 56.12\%
accuracy and 56.04\% Macro-F1. At 30 dB, the corresponding values
are approximately 55.88\% and 55.57\%.

This indicates that the major performance improvement occurs as the
signal moves from extremely noisy conditions toward moderate SNR,
after which performance becomes comparatively stable.

\section{Raw-IQ CNN versus Semantic Representations}

To establish a direct baseline comparison, we compare a conventional
CNN operating on the raw IQ representation with semantic classifiers
operating on 64D and 16D representations.

The raw-IQ CNN receives the original 2048-dimensional representation,
whereas the semantic models operate on substantially smaller latent
representations.

\begin{figure}[t]
    \centering
    \includegraphics[width=\linewidth]
    {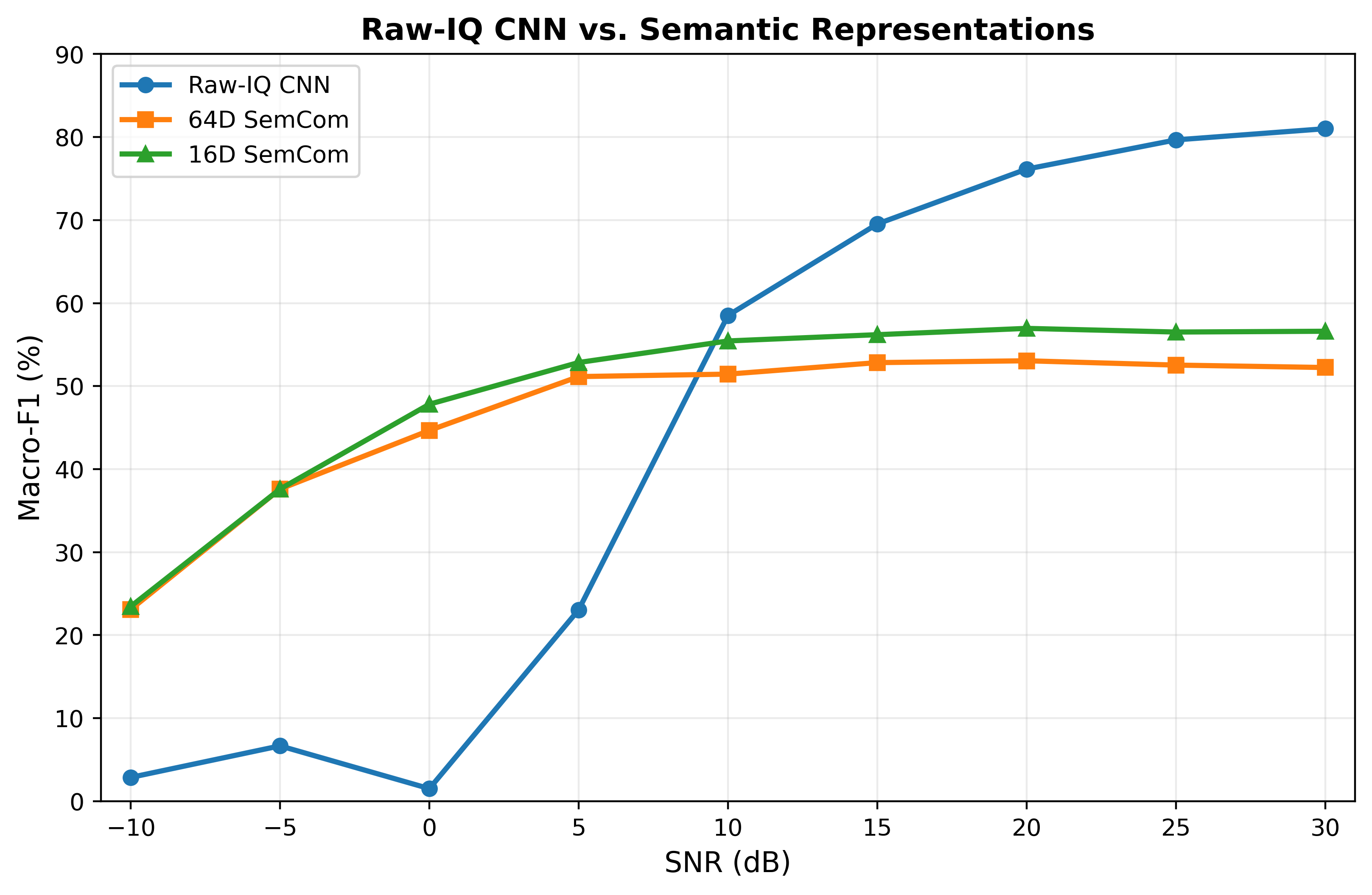}
    \caption{
    Macro-F1 comparison between a raw-IQ CNN, a 64D semantic
    representation, and the proposed 16D semantic representation
    across SNR. The semantic representations provide substantially
    stronger performance in the low-SNR regime, while the raw-IQ CNN
    becomes stronger at high SNR.
    }
    \label{fig:cnncomparison}
\end{figure}

The comparison reveals a clear difference between the operating
regimes of the models. At very low SNR, both semantic representations
provide considerably higher Macro-F1 than the raw-IQ CNN.

The 16D representation also remains competitive with the 64D
representation despite using only one quarter of the latent
dimensions. At higher SNR, the raw-IQ CNN benefits from access to
the full signal representation and eventually exceeds the semantic
models.

This comparison highlights the central trade-off of the proposed
approach: semantic compression substantially reduces representation
size while maintaining useful classification performance and
improving robustness in the low-SNR regime.

\section{Representation Stability under Noise}

Classification performance alone does not fully characterize
representation quality. We therefore examine how the learned 16D
representation changes as the channel becomes noisier.

The noisy representation is compared with a high-SNR reference.
Cosine similarity and distance-based measures quantify the degree of
noise-induced displacement in the representation space.

\begin{figure}[t]
    \centering
    \includegraphics[width=\linewidth]
    {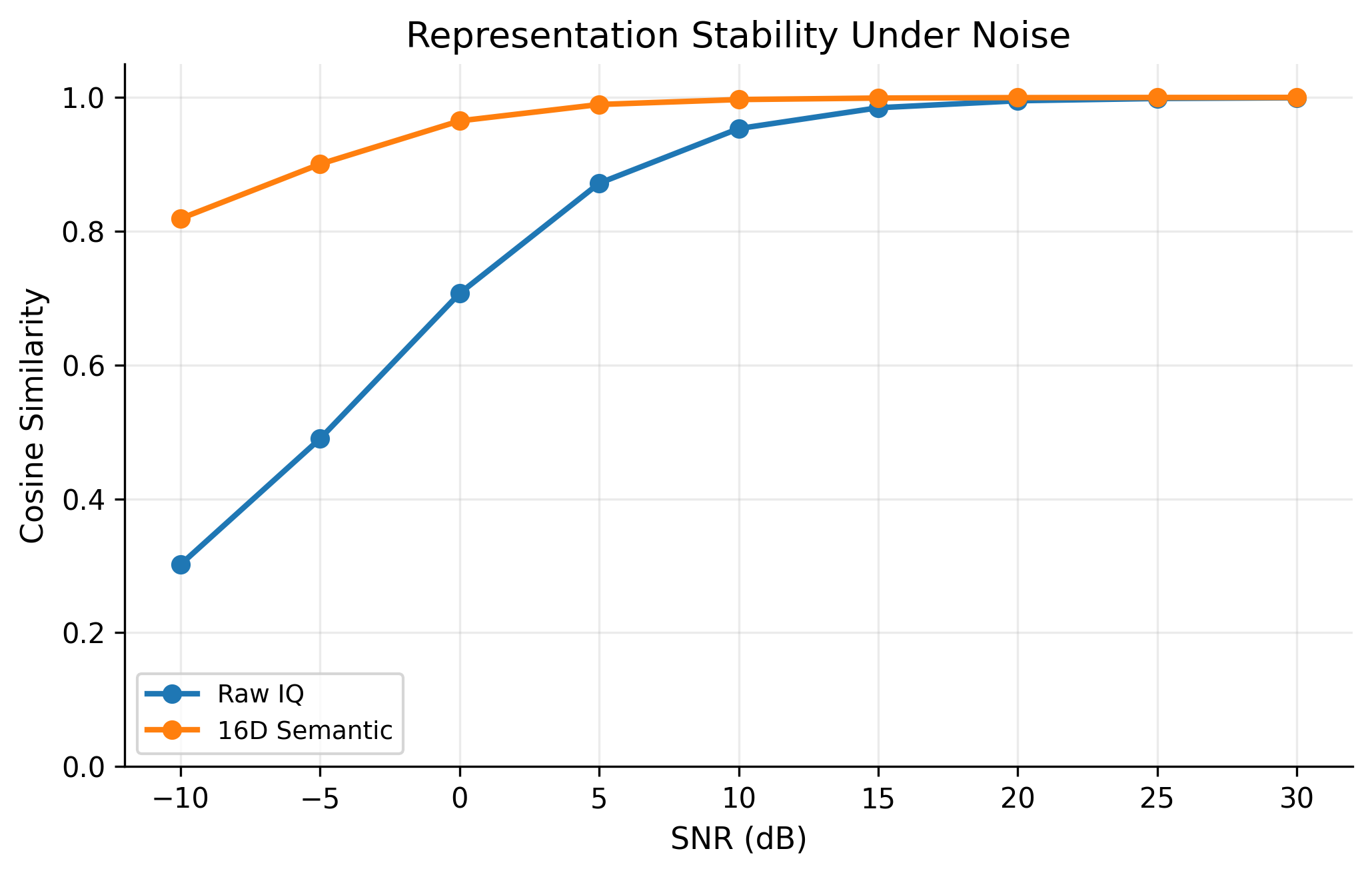}
    \caption{
    Representation stability under SNR variation. The 16D semantic
    representation remains substantially more stable than the raw
    IQ representation, particularly in the low-SNR regime.
    }
    \label{fig:stability}
\end{figure}

At $-10$ dB, the raw IQ representation has a cosine similarity of
approximately 0.3018 with the high-SNR reference, whereas the
16D semantic representation achieves approximately 0.8189.

At higher SNR, both representations become increasingly close to
their reference conditions. The difference is most pronounced under
severe noise, indicating that the semantic encoder suppresses a
substantial portion of noise-induced variation.

\section{Per-Class Robustness}

Representation robustness is not necessarily uniform across
modulation classes. We therefore analyze performance individually
for all 11 classes.

\begin{figure}[t]
    \centering
    \includegraphics[width=\linewidth]
    {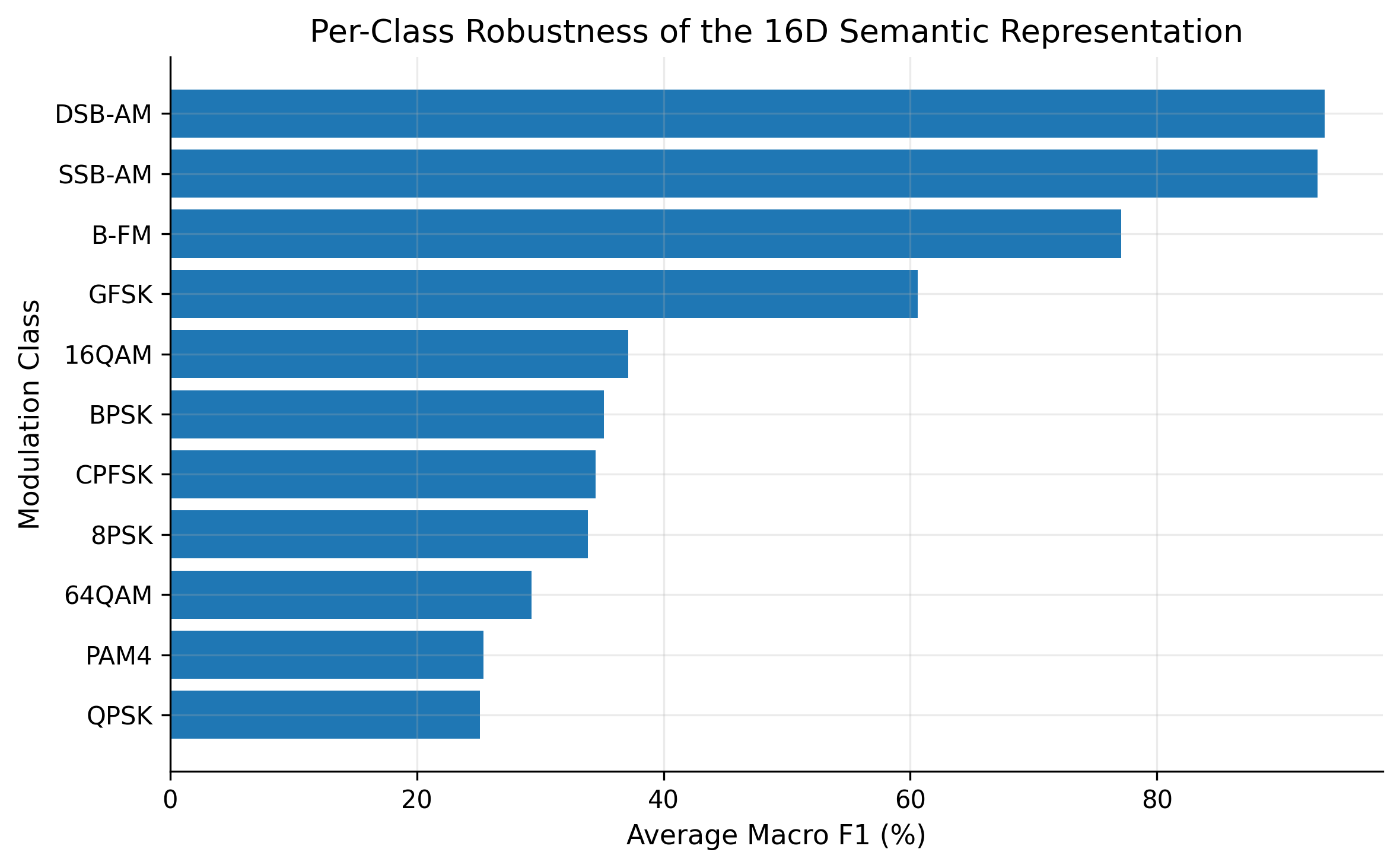}
    \caption{
    Per-class robustness of the selected 16D semantic
    representation. Performance varies substantially across
    modulation classes.
    }
    \label{fig:perclass}
\end{figure}

DSB-AM provides the highest average F1 at 93.60\%, followed by
SSB-AM at 93.01\%. QPSK has the lowest average F1 at 25.11\%.

At low SNR, SSB-AM remains particularly robust, reaching 83.26\%
F1, whereas 64QAM reaches only 18.31\%.

\begin{table}[t]
\caption{Per-class robustness of the 16D representation.}
\label{tab:perclass}
\centering
\begin{tabular}{lcccc}
\toprule
Class & Avg. Acc. & Avg. F1 & Low F1 & High F1 \\
\midrule
16QAM  & 35.41 & 37.15 & 25.79 & 42.48 \\
64QAM  & 29.63 & 29.28 & 18.31 & 34.57 \\
8PSK   & 37.63 & 33.85 & 25.66 & 37.84 \\
B-FM   & 79.41 & 77.09 & 55.80 & 88.37 \\
BPSK   & 37.33 & 35.16 & 25.15 & 40.48 \\
CPFSK  & 32.89 & 34.47 & 22.44 & 41.21 \\
DSB-AM & 91.41 & 93.60 & 81.25 & 100.00 \\
GFSK   & 57.48 & 60.59 & 38.22 & 71.91 \\
PAM4   & 26.07 & 25.40 & 22.29 & 26.80 \\
QPSK   & 24.74 & 25.11 & 19.60 & 27.75 \\
SSB-AM & 90.52 & 93.01 & 83.26 & 98.15 \\
\bottomrule
\end{tabular}
\end{table}

These results show that semantic compression does not eliminate
class-dependent ambiguity. Instead, it preserves strong performance
for some modulation families while leaving challenging classes that
require further investigation.

\section{Multi-Seed Validation}

To evaluate statistical stability, the main robustness experiment was
repeated over five random seeds.

\begin{figure}[t]
    \centering
    \includegraphics[width=\linewidth]
    {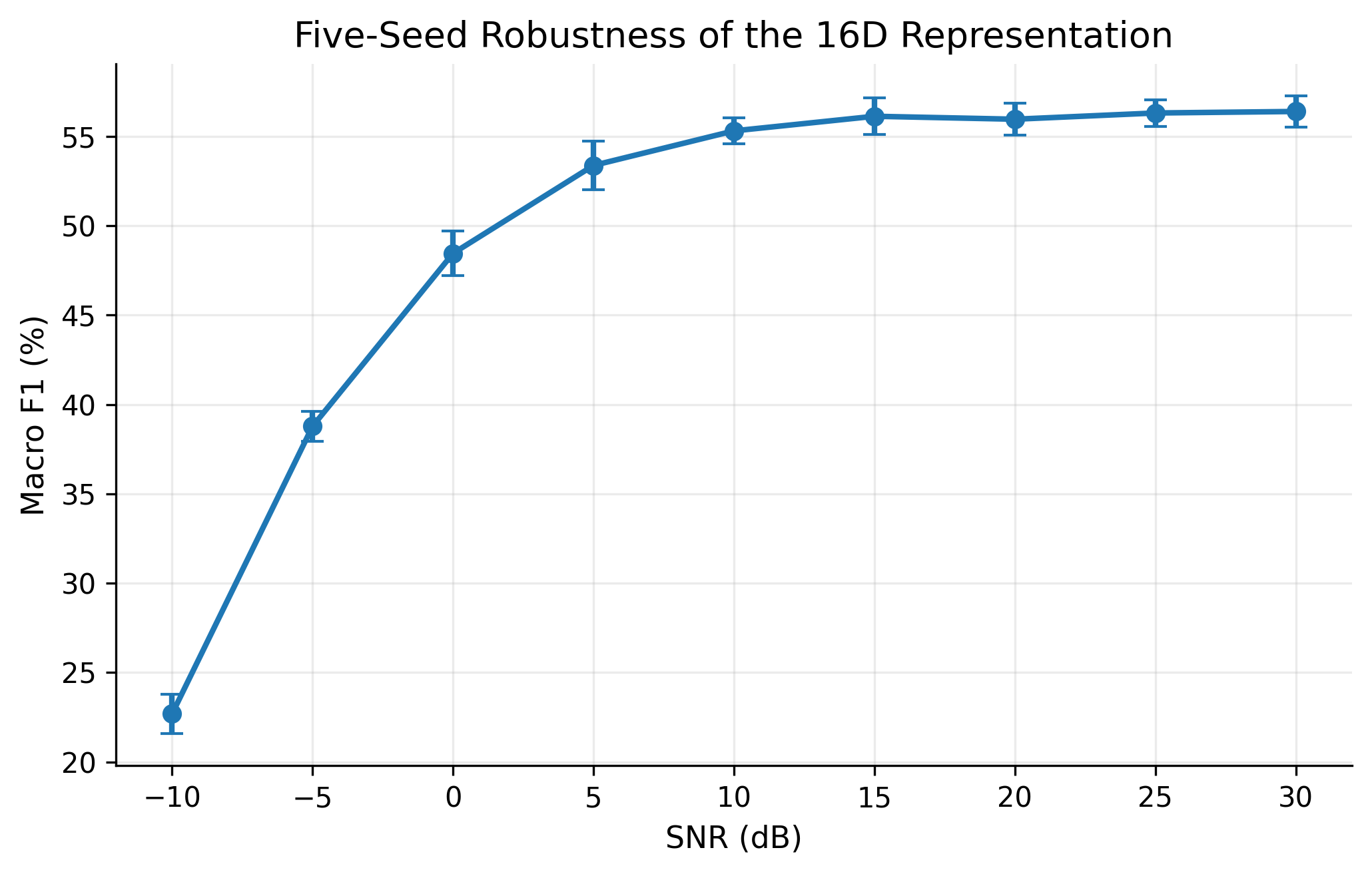}
    \caption{
    Five-seed robustness of the 16D semantic representation across
    SNR. Error bars or shaded variation represent seed-level
    variability.
    }
    \label{fig:multiseed}
\end{figure}

Across all evaluated SNR values and seeds, the average accuracy is
$49.22\pm0.52\%$, while the average Macro-F1 is
$49.27\pm0.53\%$.

The corresponding low-SNR Macro-F1 is $40.82\pm0.63\%$, compared
with $56.02\pm0.73\%$ at high SNR.

The seed-to-seed standard deviations of accuracy and Macro-F1 are
0.523\% and 0.533\%, respectively. These relatively small variations
indicate that the principal performance trend is not dependent on a
single random initialization.

\begin{table}[t]
\caption{Five-seed robustness results. Values are mean $\pm$
standard deviation.}
\label{tab:multiseed}
\centering
\begin{tabular}{rcc}
\toprule
SNR & Accuracy (\%) & Macro-F1 (\%) \\
\midrule
$-10$ & $21.62\pm0.96$ & $22.68\pm1.11$ \\
$-5$  & $37.58\pm0.86$ & $38.78\pm0.85$ \\
0     & $48.00\pm1.38$ & $48.46\pm1.25$ \\
5     & $53.55\pm1.31$ & $53.38\pm1.35$ \\
10    & $55.66\pm0.63$ & $55.31\pm0.73$ \\
15    & $56.58\pm1.03$ & $56.13\pm1.01$ \\
20    & $56.39\pm0.83$ & $55.96\pm0.88$ \\
25    & $56.78\pm0.68$ & $56.31\pm0.74$ \\
30    & $56.85\pm0.84$ & $56.39\pm0.88$ \\
\bottomrule
\end{tabular}
\end{table}

\section{Efficiency and Compression Analysis}

The selected representation substantially reduces the amount of
information that must be stored or communicated.

\begin{figure}[t]
    \centering
    \includegraphics[width=\linewidth]
    {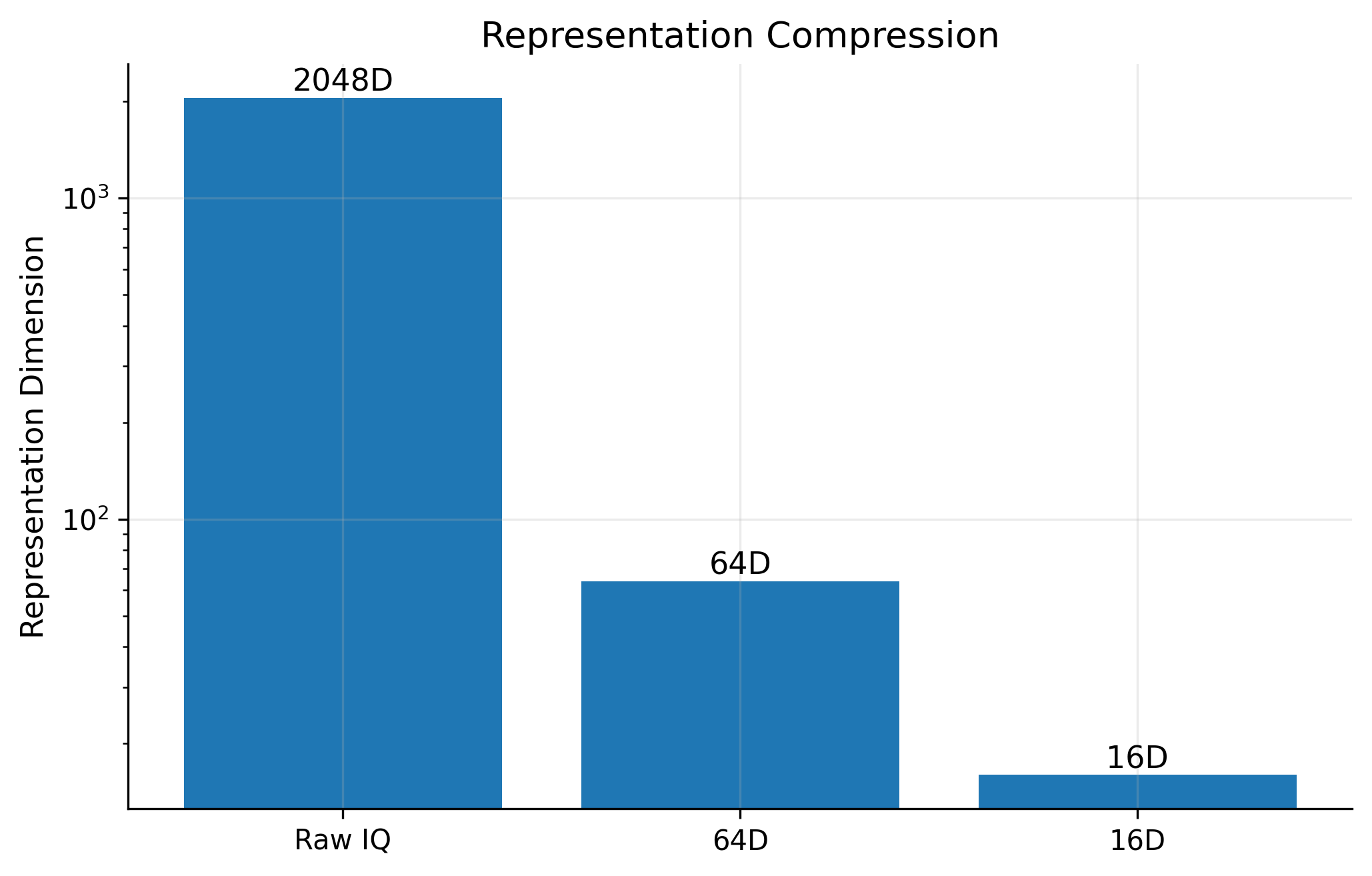}
    \caption{
    Representation efficiency of the selected 16D semantic
    representation compared with the original 2048D representation.
    }
    \label{fig:efficiency}
\end{figure}

The representation size decreases from 6.4453 MB to 0.0504 MB for
the reported comparison, corresponding to a $128\times$ memory
reduction.

The measured encoder latency is 0.002009 ms per sample, corresponding
to approximately 497831.88 samples/s under the reported evaluation
setup.

\begin{table}[t]
\caption{Representation and encoder efficiency.}
\label{tab:efficiency}
\centering
\begin{tabular}{lc}
\toprule
Metric & Value \\
\midrule
Original representation & 2048D \\
Semantic representation & 16D \\
Compression ratio & $128\times$ \\
Dimension reduction & 99.22\% \\
Encoder parameters & 1,116,816 \\
Raw representation memory & 6.4453 MB \\
16D representation memory & 0.0504 MB \\
Memory reduction & $128\times$ \\
Encoder latency/sample & 0.002009 ms \\
Encoder throughput & 497831.88 samples/s \\
Encoder MACs/sample & 1,116,160 \\
\bottomrule
\end{tabular}
\end{table}

These measurements demonstrate that the compression is not merely a
reduction in abstract dimensionality: it directly reduces the size
of the latent information that must be stored or transmitted.

\section{Discussion}

The experiments reveal several important properties of learned
semantic representations for AMC.

First, compression and classification performance do not follow a
simple monotonic relationship. Although larger latent dimensions
provide greater representational capacity, the 16D configuration
performs strongly compared with the other evaluated configurations.
This suggests that an appropriate information bottleneck can be
beneficial rather than merely restrictive.

Second, the representation exhibits substantially greater geometric
stability than raw IQ under low-SNR perturbations. At $-10$ dB,
raw IQ has a cosine similarity of 0.3018 with the reference, whereas
the 16D representation retains a similarity of 0.8189.

Third, the CNN comparison shows that semantic representations and
full raw-IQ models have different robustness regimes. The semantic
representations are particularly competitive in the low-SNR regime,
while the raw-IQ CNN benefits from its access to the complete
high-dimensional signal at higher SNR.

Fourth, robustness is class dependent. DSB-AM and SSB-AM remain
highly robust, whereas QPSK, PAM4, and 64QAM remain substantially
more difficult.

The results also suggest that representation robustness and
classification accuracy should be treated as complementary
quantities. A representation may remain geometrically stable while
the classifier still confuses classes that are close in semantic
space. This motivates future work on explicitly structuring the
latent space according to class separation and channel robustness.

\section{Limitations}

Several limitations should be considered when interpreting the
results.

First, the study evaluates a fixed set of modulation classes and a
specific experimental dataset. Generalization to additional
modulation families and independently collected RF datasets remains
to be established.

Second, the representation robustness analysis uses a 30 dB
observation as the clean reference. Although this provides a
consistent reference for the current study, it does not represent
an ideal noiseless waveform.

Third, the efficiency measurements are hardware dependent. The
reported latency and throughput therefore describe the evaluation
environment rather than an implementation-independent guarantee.

Finally, the selected 16D representation is optimal among the
evaluated dimensions, not necessarily globally optimal. Intermediate
dimensions or alternative encoder architectures could potentially
produce better accuracy--compression trade-offs.

\section{Conclusion}

This work investigated learned semantic compression for robust
automatic modulation classification. Starting from a
2048-dimensional raw IQ representation, we systematically evaluated
semantic dimensions of 4, 8, 16, 32, and 64 dimensions.

The dimension ablation identified 16D as the strongest configuration
among the evaluated dimensions. The resulting representation provides
a $128\times$ compression ratio and reduces representation
dimensionality by 99.22\%.

Beyond classification performance, we examined representation
stability under SNR variation. The 16D representation remains
substantially more similar to a clean reference than raw IQ at low
SNR. At $-10$ dB, the semantic representation achieves cosine
similarity 0.8189 compared with 0.3018 for raw IQ.

The CNN comparison further demonstrates that the semantic
representations provide a useful low-SNR operating regime while
requiring substantially fewer representation dimensions.

Per-class analysis shows that robustness is not uniform across
modulation types, while five-seed experiments demonstrate that the
principal performance trends are statistically stable. Finally, the
efficiency analysis confirms that the compact representation can
substantially reduce representation memory and communication burden.

Overall, the results support learned semantic compression as a
promising approach for resource-constrained and noise-robust AMC.
Future work should investigate larger and independently collected RF
datasets, more diverse channel impairments, alternative latent
dimensions, and objectives that explicitly optimize the trade-off
between semantic compression, class separation, and robustness.

\subsection*{AI Use Statement}

Generative AI tools were used during the preparation of this
manuscript for assistance with drafting, organization, editing, and
code-related reasoning. All experimental results reported in this
manuscript were generated from the authors' experimental pipeline and
were reviewed against the recorded experiment outputs. The authors
take responsibility for the final manuscript, including all
scientific claims, experimental descriptions, and interpretations.

\subsection*{Ethics Statement}

This work does not involve human subjects, personal data, or human
experimentation. The study focuses on machine learning
representations for automatic modulation classification. The authors
are responsible for ensuring that the reported experiments and
conclusions accurately reflect the underlying experimental results.

\subsection*{Reproducibility Statement}

The experimental pipeline is organized into dimension ablation,
representation stability, raw-IQ comparison, per-class robustness,
multi-seed validation, and efficiency experiments.

The reported experiments use a fixed 2048-dimensional input
representation, semantic dimensions of 4, 8, 16, 32, and 64
dimensions, nine SNR values from $-10$ to $30$ dB, and 11 modulation
classes.

The five-seed experiment reports mean and standard deviation to
quantify sensitivity to random initialization. Experimental outputs
include CSV and text summaries for the individual experiments.

\subsubsection*{Acknowledgments}

We thank the members of our research environment for discussions and
technical feedback that contributed to this work.

\bibliographystyle{iclr2027_conference}
\bibliography{references}

\newpage
\appendix

\section{Additional Experimental Details}

\subsection{Semantic Dimensions}

The evaluated semantic dimensions were

\[
\{4,8,16,32,64\}.
\]

Their corresponding compression ratios relative to the
2048-dimensional input are:

\begin{table}[h]
\centering
\caption{Semantic dimensions and compression ratios.}
\label{tab:appendixdimensions}
\begin{tabular}{cc}
\toprule
Dimension & Compression ratio \\
\midrule
4D  & $512\times$ \\
8D  & $256\times$ \\
16D & $128\times$ \\
32D & $64\times$ \\
64D & $32\times$ \\
\bottomrule
\end{tabular}
\end{table}

\subsection{SNR Evaluation}

The main robustness evaluation uses

\[
-10,-5,0,5,10,15,20,25,30\ \mathrm{dB}.
\]

The 30 dB condition is used as the high-quality reference condition
for representation stability analysis.

\subsection{Per-Class Results}

The 11 modulation classes evaluated in the study are:

\begin{enumerate}
    \item 16QAM (16-Quadrature Amplitude Modulation)
    \item 64QAM (64-Quadrature Amplitude Modulation)
    \item 8PSK (8-Phase Shift Keying)
    \item B-FM (Broadcast Frequency Modulation)
    \item BPSK (Binary Phase Shift Keying)
    \item CPFSK (Continuous-Phase Frequency Shift Keying)
    \item DSB-AM (Double-Sideband Amplitude Modulation)
    \item GFSK (Gaussian Frequency Shift Keying)
    \item PAM4 (4-Level Pulse Amplitude Modulation)
    \item QPSK (Quadrature Phase Shift Keying)
    \item SSB-AM (Single-Sideband Amplitude Modulation)
\end{enumerate}

\subsection{Five-Seed Evaluation}

The final multi-seed evaluation produced:

\begin{table}[H]
\centering\small
\caption{Summary of five-seed validation.}
\label{tab:appendixmultiseed}
\begin{tabular}{lc}
\toprule
Metric & Result \\
\midrule
Average Accuracy & $49.22\pm0.52\%$ \\
Average Macro-F1 & $49.27\pm0.53\%$ \\
Low-SNR Macro-F1 & $40.82\pm0.63\%$ \\
High-SNR Macro-F1 & $56.02\pm0.73\%$ \\
Seed Accuracy STD & 0.523\% \\
Seed F1 STD & 0.533\% \\
\bottomrule
\end{tabular}
\end{table}

\subsection{Efficiency Measurements}

The selected 16D encoder contains 1,116,816 parameters. The original
2048D representation requires 6.4453 MB in the reported comparison,
whereas the 16D representation requires 0.0504 MB.

The measured encoder latency is 0.002009 ms/sample, with a reported
throughput of 497831.88 samples/s and 1,116,160 MACs/sample.

\end{document}